\documentclass{aastex62}
\usepackage{amsmath}

\graphicspath{{./}{figures/}}

\received{}
\revised{}
\accepted{\today}
\submitjournal{ApJ}

\shorttitle{Long GRB cosmic history}
\shortauthors{Ghirlanda \& Salvaterra}

\newcommand{\ep}{$E_{\rm p}$}
\newcommand{\eiso}{$E_{\rm iso}$}
\newcommand{\liso}{$L_{\rm iso}$}
\newcommand{\ama}{$E_{\rm p}-E_{\rm iso}$}
\newcommand{\yone}{$E_{\rm p}-L_{\rm iso}$}
\newcommand{\jet}{$\theta_{\rm jet}$}
\newcommand{\view}{$\theta_{\rm v}$}
\newcommand{\G}{$\Gamma_0$}
\newcommand{\swift}{{\it Swift }}
\newcommand{\fermi}{{\it Fermi }}

\begin{document}

\title{THE COSMIC HISTORY OF LONG GAMMA RAY BURSTS}

\correspondingauthor{Giancarlo Ghirlanda}
\email{giancarlo.ghirlanda@inaf.it}

\author[0000-0001-5876-9259]{Giancarlo Ghirlanda}
\affiliation{Istituto Nazionale di Astrofisica (INAF) \\
Osservatorio astronomico di Brera
Via E. Bianchi 46, I-23807, Merate, Italy}
\affiliation{INFN – Sezione di Milano-Bicocca, Piazza della Scienza 3, 20126 Milano, Italy}

\author[0000-0002-9393-8078]{Ruben Salvaterra}
\affiliation{Istituto Nazionale di Astrofisica (INAF) \\
Istituto di Astrofisica Spaziale e Fisica cosmica,
via Alfonso Corti 12, 20133 Milano, Italy}


\begin{abstract}
The cosmic formation rate of long Gamma Ray Bursts (LGRBs) encodes the evolution, across cosmic times, of their progenitors' properties and of their environment. The LGRB formation rate and the luminosity function, with its redshift evolution, are derived by reproducing the largest set of observations collected in the last four decades, namely the observer--frame prompt emission properties of GRB samples detected by the \fermi\ and Compton Gamma Ray Observatory (CGRO) satellites and the redshift, luminosity and energy distributions of flux--limited, redshift complete, samples of GRBs detected by \swift. The model that best reproduces all these constraints consists of a GRB formation rate increasing with redshift $\propto (1+z)^{3.2}$, i.e. steeper than the star formation rate, up to $z\sim3$ followed by a decrease $\propto(1+z)^{-3}$. On top of this, our model predicts also a moderate evolution of the characteristic luminosity function break $\propto(1+z)^{0.6}$. Models with only luminosity or rate evolution are excluded at $>5\sigma$ significance. The cosmic rate evolution of LGRBs is interpreted as their preference to occur in environments with metallicity $12+\log(\rm O/H)<8.6$, consistently with theoretical models and host galaxy observations. The LGRB rate at $z=0$, accounting for their collimation, is $\rho_0=79^{+57}_{-33}$ Gpc$^{-3}$ yr$^{-1}$ (68\% confidence interval). This corresponds to $\sim$1\% of broad--line Ibc supernovae producing a successful jet in the local Universe. This fraction increases up to $\sim$7\% at $z\ge3$. Finally, we estimate that at least $\approx0.2-0.7$ yr$^{-1}$ of \swift\ and \fermi\ detected bursts at $z<0.5$ are jets observed slightly off--axis.  

\end{abstract}



\section{Introduction} \label{sec:intro}
Distinctive features of Long durantion Gamma Ray Bursts (LGRBs) such as their detection up to high redshifts (z=8.2--9.2 - \citealt{Salvaterra2009-ha}, \citealt{Tanvir2009-lp}, \citealt{Cucchiara2011-wb}), their association with faint blue star forming galaxies, their positional coincidence with the brightest UV regions within their hosts \citep{Fruchter2006-ni} and their association with core--collapase supernovae \citep{Hjorth2011-bt} point to a massive star progenitor thus establishing a direct link with the formation and relatively rapid evolution of massive stars across cosmic times. The study of increasingly larger samples of hosts revealed  the preference of LGRBs to form in sub--solar metallicity environments \citep{Vergani2015-rr,Perley2016-sd,Palmerio2019-zp}. This is in agreement with progenitors models \citep{MacFadyen1999-pd} requiring a low metallicity to keep high specific angular momentum at the time of the core--collapse in order to produce the GRB jet \citep{Yoon2006-ic}. Therefore, the physical conditions for the production of a relativistic jet determine the intrinsic properties of the LGRB population such as their characteristic luminosities and cosmic rate. 

GRB population studies aims at reconstructing the population luminosity and formation rate functions. The free parameters of these functions are often constrained by reproducing a set of rest frame and observer frame properties \citep{Daigne2006-dy,Wanderman2010-cb,Robertson2012} of detected GRB samples 
and by properly accounting for sample incompleteness \citep{Salvaterra2012-uc,Ghirlanda2015-zv,Palmerio2020-sh}. 


In this work we build a GRB population model (\S\ref{sec:model}) and derive the cosmic formation rate of LGRBs and their luminosity function together with its evolution with redshift.  
To this aim we exploit, for the first time, the largest set of observational constraints collected by different GRB detectors in the last four decades (\S\ref{constraints}). 
We account for the presence of relativistic jets and for the orientation--dependence of the observed GRB properties. We derive the evolution of the ratio of GRB--to--cosmic--star formation rate and test its consistency with the host metallicity bias (\S\ref{sec:results}). We discuss implications of our population model in \S\ref{discussion}. Throughout the paper we assume standard flat cosmology $\Omega_M=0.3$, $h=0.7$.

\begin{figure*}[!hbt]
    \centering
    \includegraphics[scale=0.8]{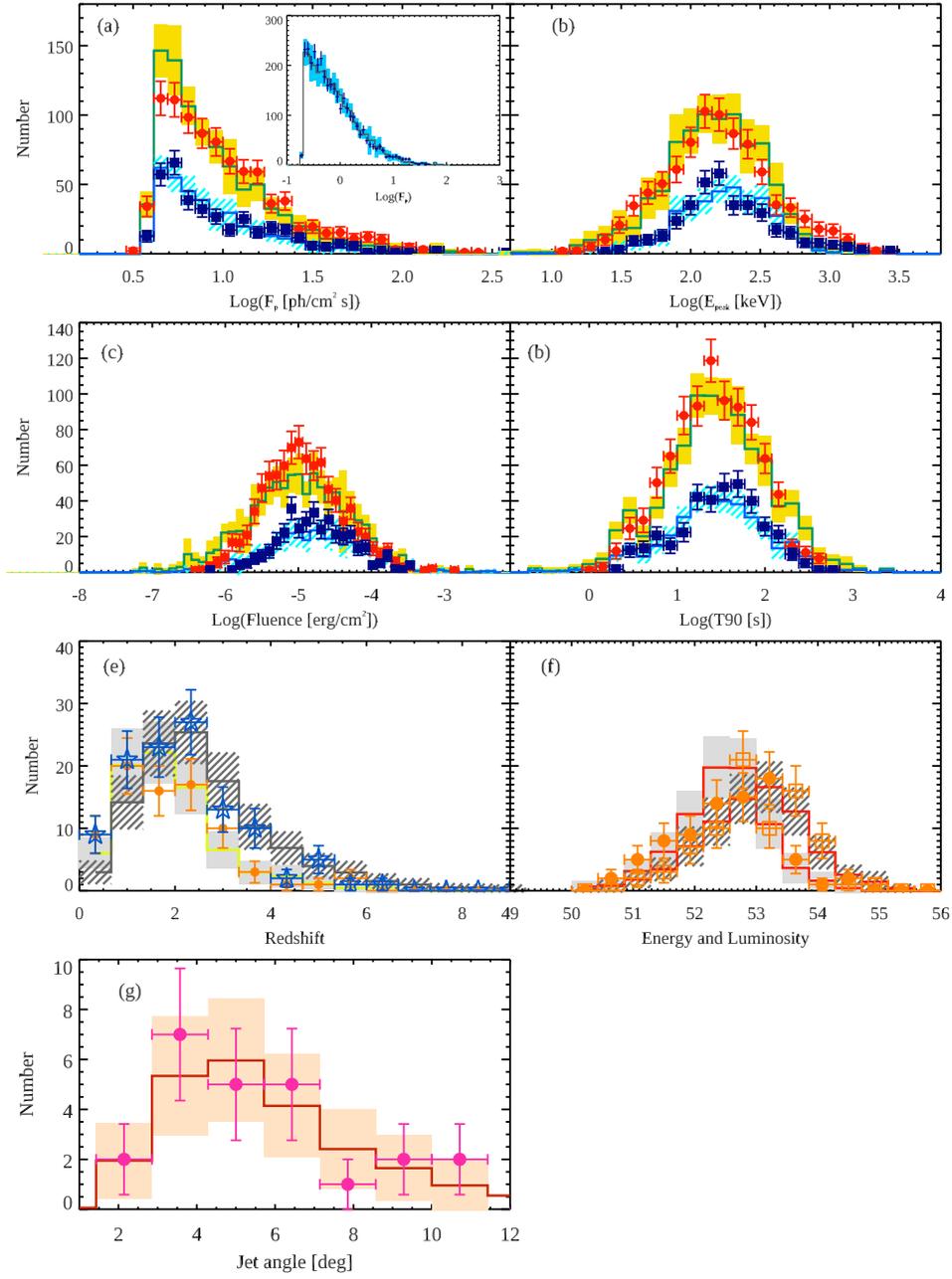}
    \vskip -1truecm
    \caption{Observational constraints of the model population. (a) Distributions of peak flux on 1.024 s timescale integrated over the 10--1000 keV energy range for \fermi/GBM  (red cicles) and CGRO/BATSE  (blue squares) GRBs with with $F_p\ge 4$ ph cm$^{-2}$ s$^{-1}$. The inset shows the peak flux (50--300 keV) of BATSE GRBs reconstructed by an off-line search \citep{Stern2001-rt} with $F_p\ge$0.3 ph cm$^{-2}$ s$^{-1}$. Panels (b), (c) and (d) show the \ep,  fluence $F$ (10-1000 keV) and  $T_{90}$ distributions (for GBM and BATSE with red and blue symbols respectively). Panel (e) shows the redshift distributions of the \swift BAT6 complete sample  (orange circles) and of  the \swift SHOALS sample (blue stars) while panel (f) shows the \liso\ and \eiso\ (open squares and filled circles respectively) distributions of the BAT6 sample. Panel (g) shows the jet opening angle distribution (circles). In each panel the solid line histogram and the shaded (dashed) region show the best model and its 1$\sigma$ uncertainty respectively obtained by the MCMC method described in \$\ref{sec:model}. The model uncertainty is obtained by sampling the parameter posterior distributions of the free parameters (shown in Fig.\ref{fig:corner}) and including Poisson uncertainty. 
   }
    \label{fig:constr}
\end{figure*}
\section{The model}\label{sec:model}
We build a synthetic LGRB population model based on two distribution functions. The GRB formation rate per unit comoving volume $\rho(z)$ is modelled with the parametric function \citep{Cole2001-tx,Madau2014-wg}: 
    \begin{equation}
        \rho(z)=\rho_0\frac{(1+z)^{p_{z,1}}}{1+\left(\frac{1+z}{p_{z,2}}\right)^{p_{z,3}}}
        \label{grbfr}
    \end{equation}
    where $\rho_0$, in Gpc$^{-3}$ yr$^{-1}$, is the LGRB local density rate. The choice of this functional form is  motivated by its relatively small number of free parameters and, for the ease of comparison, by its systematic use to fit the Cosmic Star Formation Rate (CSFR - \citealt{Madau2014-wg}; \citealt{Madau2017-hz}; \citealt{Hopkins2006}). Flat priors are considered for the the free parameters $(p_{z,1},p_{z,2},p_{z,3})$ which can vary over wide ranges including also the values corresponding to the CSFR. This ensures that we can recover, if present, any significant deviation of the GRB formation rate from the CSFR. 

The luminosity function of LGRBs is often described in the literature by a broken power--law (e.g. \citealt{Salvaterra2009-ha}; \citealt{Salvaterra2012-uc}):
\begin{equation}
\Phi(L_{\rm iso},z) \propto
\begin{cases}
L_{\rm iso}^{-a_{1}} & \text{if}\,\,\, L_{\rm iso}\le L_{\rm iso,b}=L_{\rm iso,0}(1+z)^{\delta}\\
L_{\rm iso}^{-a_{2}} &\text{if}\,\,\, L_{\rm iso}> L_{\rm iso,b}=L_{\rm iso,0}(1+z)^{\delta}
\end{cases}\label{LF}
\end{equation}
where $a_1$ and $a_2$ are the indices of the power laws respectively below and above the break luminosity $L_{\rm iso, b}$  which is left free to evolve with redshift $\propto (1+z)^{\delta}$. Therefore, $L_{\rm iso,0}$ represents the break of the luminosity function at $z=0$. $\Phi(L_{\rm iso},z)$ is normalised to its integral and it is defined for $L_{\rm iso}\ge 10^{47}$ erg/s. As alternative working hypothesis sometimes considered in the literature is a Schechter \citep{Schechter1976} luminosity function. We discuss in \S\ref{A3} the results obtained with under this different assumption. 
In summary, our model allows for the evolution of the GRB population both in terms of rate density (Eq\ref{grbfr}) and luminosity (Eq.\ref{LF}). 

LGRBs follow the $E_{\rm p}-L_{\rm iso}$ \citep{Yonetoku2004-zv} correlation, where \ep\ is the peak energy of the $\nu F_{\nu}$ spectrum, and the \ama\ correlation \citep{Amati2002-dq}, linking the peak energy 
with the isotropic energy \eiso. It has been proved \citep{Nava2012-ny, Dai2021} that these correlations do not  evolve with redshift. Following \citealt{Ghirlanda2015-zv}, in our simulation we sample Eq.\ref{grbfr} extracting a redshift and independently we assign an \ep\ value sampling a broken powerlaw. 
The values of \liso\ and \eiso\ corresponding to  \ep\ are derived from the \yone\ and \ama\ correlations, respectively, accounting for their gaussian scatters \citep{Nava2012-ny}.  
For the ease of comparison with the literature, once we have derived the posterior parameter distributions of the model, we map, through the \yone\ correlation, the peak energy distribution function back to the luminosity function of which we provide the resulting parameter values in Tab.\ref{tab:1}. 
GRB rest frame durations are estimated as $f\cdot E_{\rm iso}/L_{\rm iso}$, where $f$ is a scaling factor sampled from a free gaussian distribution $\mathfrak{G}[f|\mu,\sigma]$.

It is known that GRBs are highly collimated (e.g. \citealt{Ghirlanda2007-ch}). Therefore, we compute their intrinsic rate $\rho(z)$ (Eq.\ref{grbfr}) by accounting for the elusive bulk of the population consisting of events with jets oriented off--side our line of sight.
We assume uniform conical jets distributed with random orientations \view\ with respect to the line of sight. Jet opening angle values \jet\ appear correlated with \eiso\ \citep{Ghirlanda2007-ch} such that we assume a free parametric correlation \jet-\eiso.  Even off-axis jets (\view$>$\jet) may be detected depending on their orientation and bulk Lorentz factor $\Gamma$, characterising the prompt emission phase. We account, for the first time within a population study of LGRBs, for the relativistic effects on the observed flux, fluence and peak energy induced by the jet orientation and beaming through the doppler factor $\delta=1/\Gamma(1-\beta\cos\theta_{\rm v})$ (see e.g. \citealt{Ghisellini2006-ty}). To this aim we sample the $\Gamma$--\eiso\ correlation reported by \citealt{Ghirlanda2018-la} to assign $\Gamma$ values to simulated events.  

The GRB prompt emission spectrum is described with the empirical  Band function \citep{Band1993} consisting of two smoothly joined power laws. To each simulated GRB we assign a low and high energy spectral index extracted from gaussian distributions $\mathfrak{G}[\alpha|\mu,\sigma]$ and $\mathfrak{G}[\beta|\mu,\sigma]$ respectively, whose parameters (e.g. \citealt{Goldstein}) are given in Tab.\ref{tab:1}. 

 The population model has 12 free parameters (Tab.\ref{tab:1}): $[\rho_0,p_{z,1},p_{z,2},p_{z,3}]$ describe the GRB cosmic formation rate (Eq.\ref{grbfr}) and $[a_1,a_2,L_{\rm iso,0},\delta]$ define the luminosity function (Eq.\ref{LF}); the remaining four parameters $(\mu,\sigma)$ and $(K,S)$ describe the stretching factor of the duration distribution and the \eiso-\jet correlation respectively.
 All other parameters reported in Tab.\ref{tab:1} are fixed to their presently known values and are derived from the corresponding references given above.   

\begin{table*}[!bt]
    \caption{Parameters of the GRB population model. Parameters constrained through the MCMC (\S\ref{sec:model}) are reported with the associated 1$\sigma$ uncertainty. Their posterior distributions are shown in Fig.\ref{fig:corner}. Parameters without errors are held constant. The four correlations listed are expressed in the form $
    \log\left(\frac{Y}{Y_0}\right)=K+S\cdot\log\left(\frac{X}{X_0}\right) $ with the parameter $\Sigma$ representing their scatter along the independent coordinate.  $E_{\rm iso}, L_{\rm iso}, E_{\rm p}, \theta_{\rm jet}$ have units of erg, erg/s, keV and rad.}
    \label{tab:1}
    \centering
    \begin{tabular}{l|lllll}
    \hline
    \hline
    $\rho(z)$     & $p_{z,1}=3.33_{-0.33}^{+0.33}$ & $p_{z,2}=3.42_{-0.28}^{+0.28}$ & $p_{z,3}=6.21_{-0.32}^{+0.38}$  & $\rho_{0}=79^{+57}_{-33}$ Gpc$^{-3}$ yr$^{-1}$ & ...\\
    $\Phi(L_{\rm iso},z)$   & $a_1=0.97_{-0.04}^{+0.05}$ & $a_2=2.21_{-0.18}^{+0.13}$ & $\log L_{\rm iso,0}=52.02_{-0.19}^{+0.22}$  & $\delta=0.64^{+0.32}_{-0.26}$  & ... \\
    $\mathfrak{G}[\log f|\mu,\sigma]$ & $\mu=3.00_{-0.26}^{+0.29}$ & $\sigma=0.23_{-0.04}^{+0.04}$ & ... & ...  & ... \\
    $\mathfrak{G}[\alpha|\mu,\sigma]$ & $\mu=-1.0$ & $\sigma=0.1$ & ... & ...  & ... \\
    $\mathfrak{G}[\beta|\mu,\sigma]$ & $\mu=-2.3$ & $\sigma=0.1$ & ... & ...  & ... \\
    
    \hline
      & $K$ & $S$ & $\Sigma$  & $Y_{0}$  & $X_{0}$ \\
   \hline
   $\log$\eiso- $\log$\jet       & $9.65_{-0.01}^{+0.01}$  & $-0.203_{-0.002}^{+0.002}$ & 0.1 & 1 & 1 \\
   $\log$\ep- $\log$\liso       & 0.27 & 0.56 & 0.18  & 406.7 & $3.38\times10^{52}$ \\
   $\log$\ep- $\log$\eiso       & -0.12 & 0.58 & 0.20  & 406.7  & $9.10\times 10^{52}$ \\
   $\log$\G- $\log$\eiso       & 2.1 & 0.3 & 0.1  & 1  & $10^{52}$ \\
   \hline
    \end{tabular}
\end{table*}

The twelve free parameters are constrained by matching a set of  observed distributions $C_{J}$ (\S\ref{constraints}) and by maximising a binned likelihood. The probability that the model distribution has $m_i$ objects in the $i$ bin of the distribution, given that there are $r_{i}$ real events in this bin, follows a Poisson distribution. The total log--likelihood can be expressed as: 
\begin{equation}
    \ln\mathcal{L}=\sum_{C_{j}}\sum_{i=1}^{n}r_{i}\ln(m_i)-m_i-\ln(r_i!)
    \label{loglike}
\end{equation}
where the internal sum is performed over the $n$ bins of each distribution
 of the constraint $C_J$ (with $J=1,...,14$). The observational constraint distributions $C_{J}$ (described in \S\ref{constraints}) are build from the observed properties of  LGRB samples detected by different instruments and are represented by different simbols in the panels of Fig.\ref{fig:constr}. 

The posterior distributions of the 12 free parameters are derived through a Markov Chain Monte Carlo (MCMC) method which samples from wide uniform prior distributions of the free parameters: $p_{z,1}\in[1,6]$, $p_{z,2}\in[0.1,6]$, $p_{z,3}\in[1,9]$, $\delta\in[0,5]$, $a_{1}\in[0.5,3]$, $a_{2}\in[2,5]$, $logL_{\rm iso,b}\in[50,53]$, $\mu_{f}\in[1.5,10]$,  $\sigma_{f}\in[0.01,6]$, $K_{E-\theta}\in[9,10.5]$, $S_{E-\theta}\in[-0.25,-0.1]$. We implement the Goodman \& Weare MCMC algorithm \citep{Goodman2010-aq}. We work with 200 walkers running over 200 steps and adopt the parallel stretch move method \citep{Foreman-Mackey2013-pw} to reduce the computation time. 

\begin{figure*}
    \hskip -1.truecm
    \includegraphics[scale=1.15]{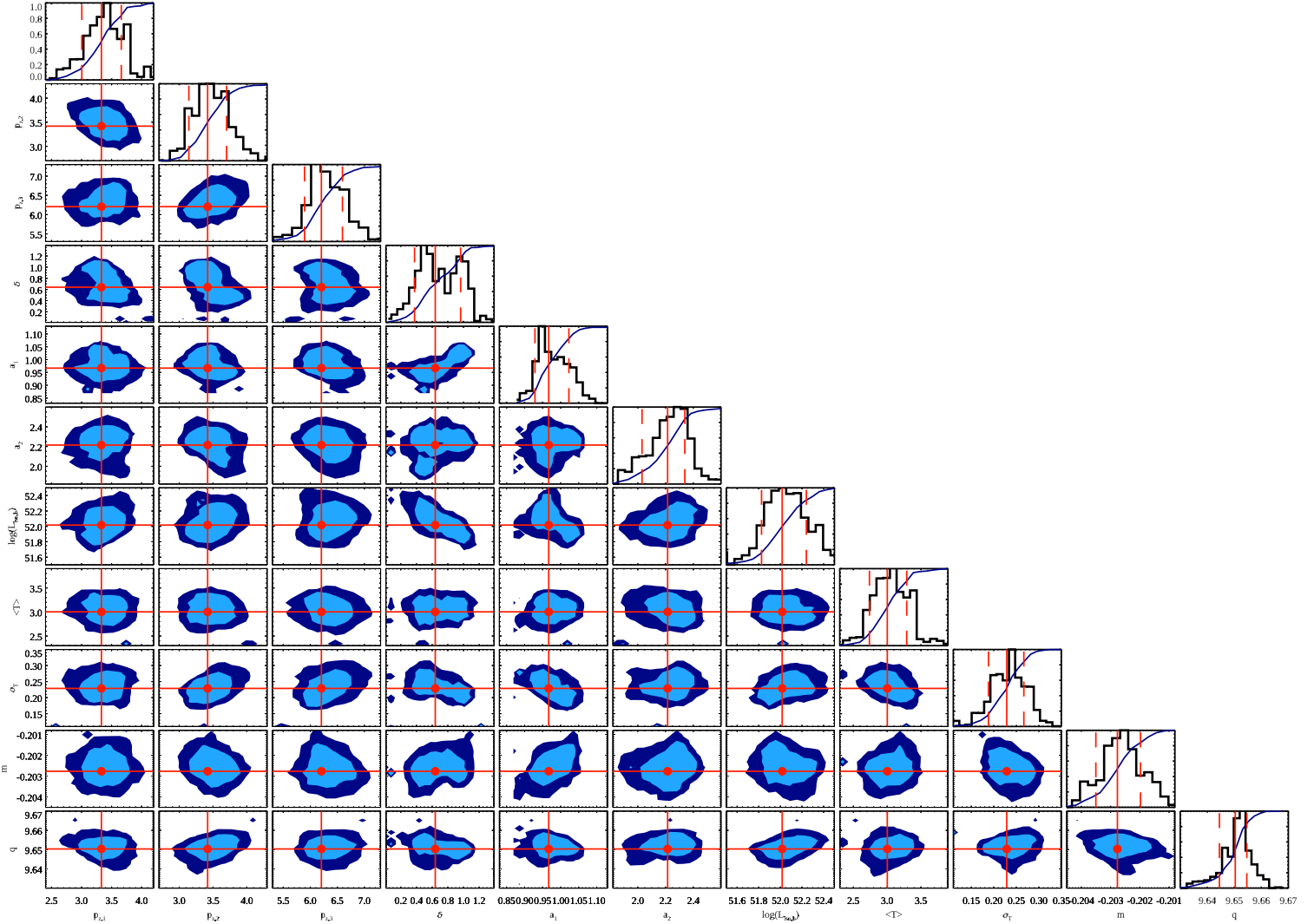}
    \caption{Corner plot showing the posterior distributions of the  12 free model parameters
    The position of the mean values of the distributions are represented by the red circles and solid red lines in the two-dimensional plots. The shaded contours for each couple of parameters represent the 68\% and 95\% contours (light and dark blue respectively). Main--diagonal plots show the histograms of the posterior values of each free parameter (with the cumulative normalized distribution shown as a thin blue line) and the solid and dashed red lines mark the central value and the 68\% confidence interval. }
    \label{fig:corner}
\end{figure*}

\section{Observational Constraints}\label{constraints}





The free parameters of the GRB population model are constrained by reproducing a set of observables' distributions provided by samples of GRBs detected by different missions.
The numerous samples of LGRBs detected by the Gamma Burst Monitor (GBM) on board the \fermi satellite 
and the Burst And Transient Source Experiment (BATSE) on board the Compton Gamma Ray Observatory (CGRO) provide the peak flux ($F_{P}$) and fluence ($F$) distributions (panels {\it a} and {\it c} in Fig.\ref{fig:constr}), the observer--frame peak energy (\ep/$1+z$) and duration ($T_{90}$) (panels {\it b} and {\it d} in Fig.\ref{fig:constr}). In order to minimize the sample incompleteness, particularly severe in the faint end of the peak flux distribution\footnote{The peak flux is reported in the \fermi online spectral catalog (https://heasarc.gsfc.nasa.gov/W3Browse/fermi/fermigbrst.html) and in the 5B BATSE catalog (https://heasarc.gsfc.nasa.gov/W3Browse/cgro/bat5bgrbsp.html) as obtained by fitting the peak spectrum with the Band function.}, we consider  LGRBs with $F_{P}\ge$4 ph cm$^{-2}$ s$^{-1}$, integrated over the 10 keV -- 1 MeV energy range for \fermi/GBM and 20 keV -- 2 MeV for GCRO/BATSE. This selection results in 904 GRBs (out of 1906) for \fermi/GBM (red symbols in Fig.\ref{fig:constr}) and 370 (out of 1529) for  CGRO/BATSE  (blue symbols in Fig.\ref{fig:constr}). The inset in Fig.\ref{fig:constr}-{\it (a)} shows the peak flux (50--300 keV) of 2810 CGRO/BATSE bursts in the Stern et al. (2001) catalog with $F_p\ge 0.3$ ph cm$^{-2}$ s$^{-1}$. 


For the redshift distribution (panel {\it e}) we consider two well selected \swift/BAT flux--limited samples for  which the large majority of bursts have a redshift measurement.
The BAT6 sample  \citep{Salvaterra2012-uc,Pescalli2016-uc} consists of {\it Swift} LGRBs with 15--150 keV peak flux $\ge$ 2.6 ph cm$^{-2}$ s$^{-1}$. 
The SHOALS sample \citep{Perley2016-sd}  comprising {\it Swift} GRBs with a 15--150 keV integrated fluence $\gtrsim 10^{-6}$ erg cm$^{-2}$.  Both these samples have 80-90\% redshift measurements either from afterglow or host spectroscopy.
 The BAT6 sample also provides the constraints on \eiso\ and \liso\ distributions (panel {\it f}). Finally, we consider the jet opening angle distribution (panel {\it g}) of GRBs collected in \citealt{Ghirlanda2007-ch}.

Further we also verified the consistency of our best model (Fig.\ref{fig:consistency} in \S\ref{A2}) with the distributions of the peak flux, fluence and duration ($T_{90}$) of the samples of LGRBs observed by the \swift/BAT\footnote{https://swift.gsfc.nasa.gov/archive/grb\_table/} and by the Gamma Ray Burst Monitor (GRBM) on board the Beppo/SAX satellite \citep{Frontera2008-el}.

\begin{figure*}[!tb]
\plotone{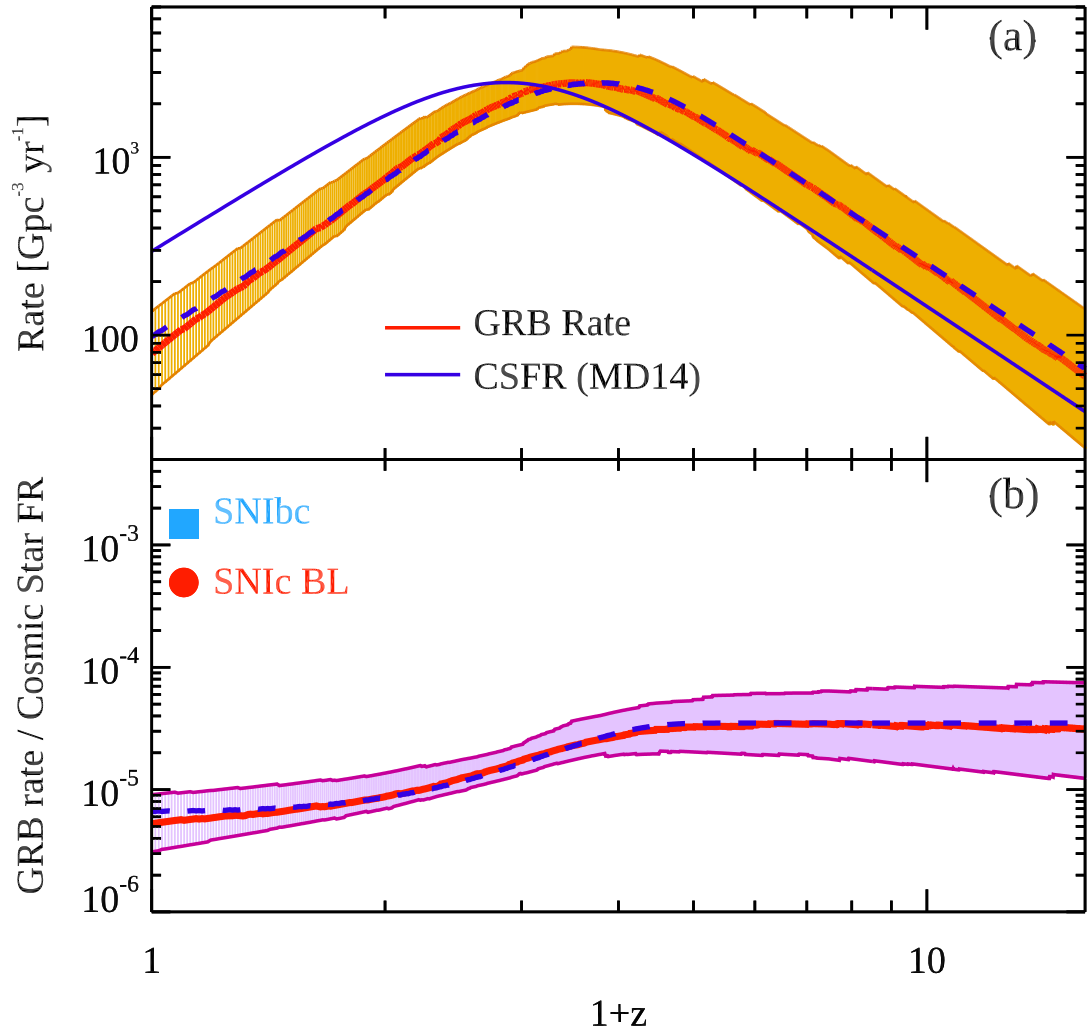}
\caption{Top panel: GRB formation rate (red solid line) and its 1$\sigma$ uncertainty (shaded region). The solid blue line shows the Cosmic Star Formation Rate (CSFR - \citealt{Madau2014-wg}). The dashed  blue lines is the CSFR modified by a metallicity threshold 12+$\log\rm(O/H)\le$8.6  (\S\ref{A3}). Blue model curves have been normalized to the peak value of the GRB formation rate. Bottom panel: GRB to CSFR ratio (red solid line with its 1$\sigma$ uncertainty). The dashed  curve, corresponding to the metallicity thresholds, is normalized to the red curve at $z\sim3$. The fraction of local supernovae (SN) of type Ibc and Ic--BL (i.e. broad line) are shown by the red circle and cyan square symbols respectively (see \citealt{Strolger2015-pa,Perley2020-sv}).  \label{fig:rate}}
\end{figure*}

\begin{figure*}[ht!]
\hskip -1cm
\includegraphics[scale=0.8]{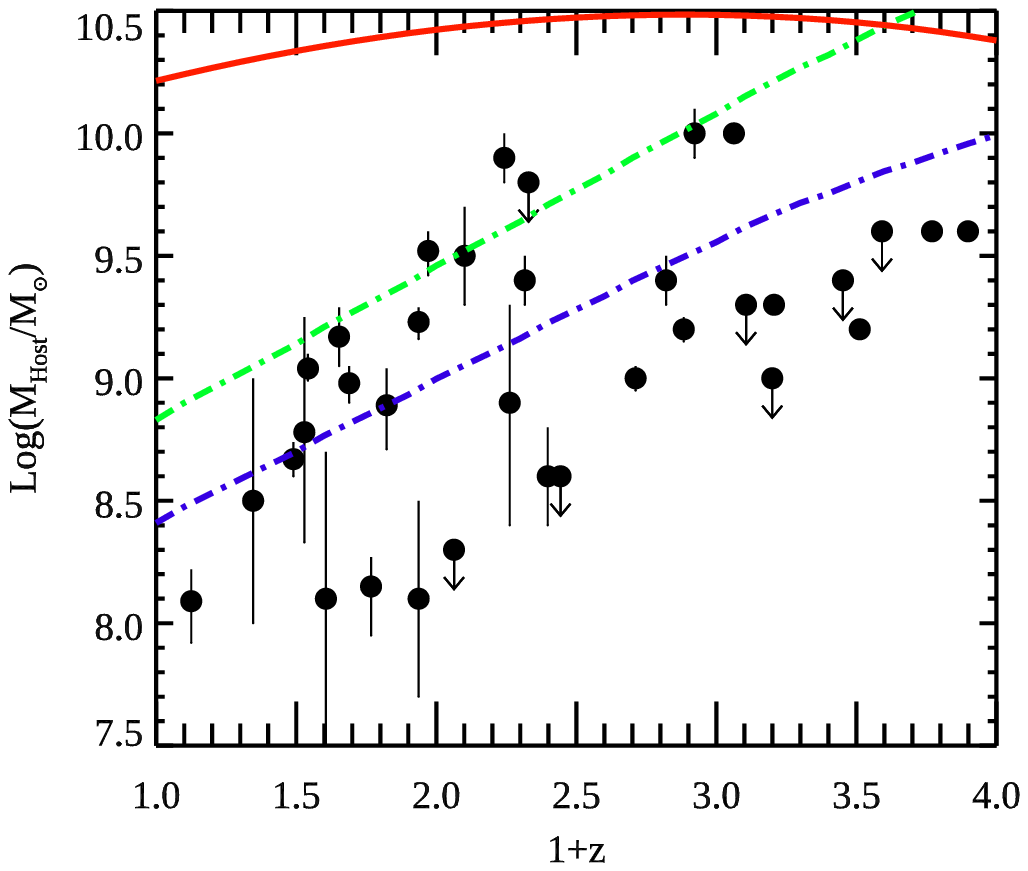}
\includegraphics[scale=0.8]{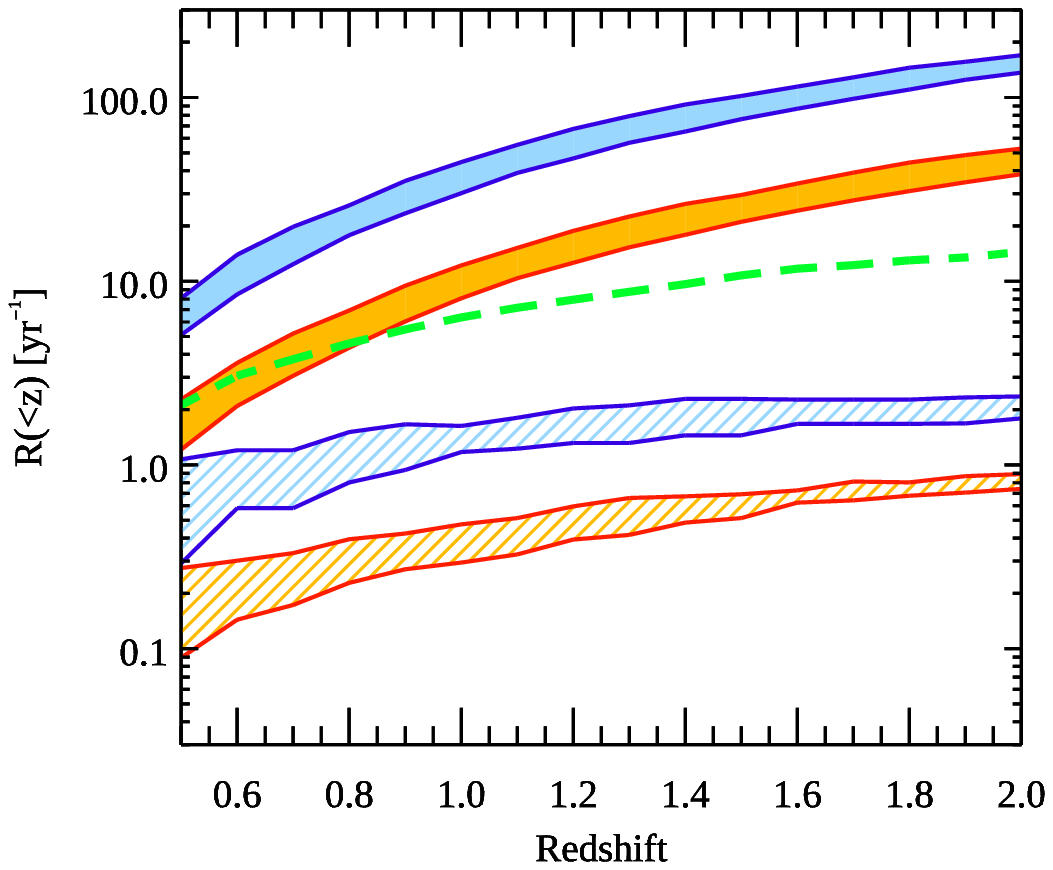}
\caption{Left panel: Mass--redshift distribution of LGRB host galaxies (black symbols - \citealt{Vergani2015-rr,Palmerio2019-zp}) compared to the average mass (blue curve) and maxiumum mass (green curve) corresponding to the  metallicity threshold 12+$\log\rm(O/H)$=8.6. The red line shows the model assuming that the GRB cosmic rate is proportional to the CSFR. Right panel: \fermi (blue) and \swift (orange) cumulative detection rates (shaded and dashed regions are 1$\sigma$ uncertainty) as a function of redshift. Dashed regions refer to off--axis events. Solid filled regions represent the total (on and off-axis). The green dashed  line  shows the rate of observed \swift GRBs with measured redshifts. This shows a reduction of the efficiency of ground based redshift measurement at $z>0.6$. } \label{fig:host}
\end{figure*}

\section{Results}\label{sec:results}

The model which best matches the set of observational constraints is shown by the solid histograms in Fig.\ref{fig:constr} (and Fig.\ref{fig:consistency}). The shaded/dashed regions around the model line represent the 68\% model uncertainty.  This is computed by resampling the joint posterior distributions of the free parameters (shown in Fig.\ref{fig:corner}) with $10^{5}$ samples and computing for each the corresponding model. For each fixed value of the independent variable the shaded region shows the 16$^{\rm th}$ to 84$^{\rm th}$ percentile of the resulting model. The posterior distributions of the best model are shown by the corner plot in Fig.\ref{fig:corner} and their central values and 1$\sigma$ uncertainty are reported in Tab.\ref{tab:1}. All the parameter values are well constrained with, on average,  $\sim10\%$ accuracy.

The luminosity function has a faint--end slope $\sim -1$ and a bright--end slope $\sim -2.2$ with a break luminosity at z=0  $L_{\rm b,0}=10^{52}$ erg s$^{-1}$. These values are consistent with those reported by previous studies  \citep{Salvaterra2012-uc,Wanderman2010-cb}. We find a mild evolution of this break which increases with redshift as $L_{b}=L_{b,0}(1+z)^{0.64}$. 


The cosmic rate of LGRBs, $\rho(z)$, is shown in Fig.\ref{fig:rate}-{\it (a)} (solid red line). The shaded orange region represents the 1$\sigma$ uncertainty as obtained by resampling the posterior distributions of the free parameters. $\rho(z)$ increases with redshift $\propto(1+z)^{3.2}$, it peaks at $z\sim3$ and declines at larger redshifts $\propto(1+z)^{-3}$. These slopes are respectively steeper and similar to those of the CSFR\footnote{For the ease of comparison, this is rescaled to the value of the GRB formation rate at its peak. } of \citealt{Madau2014-wg} (solid blue line in Fig.\ref{fig:rate}-{\it (a)}).

The local LGRB formation rate is  $\rho_0=79^{+57}_{-33}$ Gpc$^{-3}$ yr$^{-1}$ (68\% confidence interval). This is derived by normalizing the model population: we consider the $\sim$ 170$\pm$13 bright GRBs with peak flux in the 15-150 keV energy range $\ge$2.6 ph cm$^{2}$ s$^{-1}$ detected by \swift/BAT in 8 years. We assume for Swift/BAT an average duty cycle of 0.75 and the fully coded field of view of 1.4 sr. We verified that our results are unchanged if we renormalize the population to the bright GRB sample detected by \fermi/GBM, namely 902 LGRBs with P(10-1000)$\ge$ 5 ph cm$^{2}$ s$^{-1}$ detected in 11.8 years within its 8.8 sr field of view, also accounting for a 0.55 duty cycle. Since our population accounts for the random orientation of GRB jets in the sky, the derived local rate represents the intrinsic one. 

We performed the same analysis also under the assumption of a Schechter luminosity function and the results are reported in \S\ref{AS}. We find that the shape of the LGRB rate  at low redshifts is unchanged (Fig.\ref{fig:schechter}). The Schechter scenario produces a slightly shallower GRB formation rate decrease at high redshifts. This is however consistent with the CSFR given the uncertainties on high redshift measurements.

\section{Discussion and Conclusions}\label{discussion}


The LGRB efficiency as a function of redshift, defined as the ratio between the GRB rate and the cosmic star formation rate, is shown in Fig.\ref{fig:rate}--{\it b} by the solid red line.   Its 1$\sigma$ uncertainty (shaded violet region in Fig.\ref{fig:rate}) is computed by resampling the posterior distributions of the free parameters shown in Fig.\ref{fig:corner} $10^5$ times. The uncertainty For the CSFR we have considered the function of \citealt{Madau2014-wg} (solid blue line in Fig.\ref{fig:rate}-{\it a}). In the local Universe this ratio is $\sim 5\times 10^{-5} M_{\odot}^{-1}$ and it increases up to a factor of 10 at $z\sim3$. In Fig.\ref{fig:rate}-{\it b} we  show with the blue and red symbols respectively the fraction of star formation leading to Supernovae Ibc and broad line (BL) Ic \citep{Perley2020-sv} assuming the core--collapse supernova rate derived by \citealt{Strolger2015-pa}. We find that, in the local Universe, 1.3$^{+1.0}_{-0.6}$ \% of SNIc BL produce a jet and the associated GRB event. This fraction increases with redshift up to $\sim$7\% at $z>3$. 

The general trend of the GRB cosmic rate at low redshifts shows a decrease of the efficiency of producing GRBs with decreasing redshifts (Fig.\ref{fig:rate}). This result argues against the claims of a increase of the LGRB efficiency towards lower redshifts (so called low redshift excess  -- \citealt{Yu2015}; \citealt{Petrosian2015}; \citealt{Tsvetkova2017-xo}; \citealt{Lloydronning2019}, but see \citealt{Bryant2021,Pescalli2016-uc,Le2020}) obtained through non--parametric methods. 
The low redshift GRB uptick could be partly absorbed if the jet opening angle evolves with redshift \citep{Lloyd-Ronning2020-pj}.  
 


The GRB efficiency redshift evolution can be interpreted as due to the increase of the metallicity with cosmic time which prevents a larger fraction of massive stars from producing a GRB \citep{Yoon2006-ic}. In order to verify this possibility we compute the fraction of cosmic star formation in galaxies with average metallicity below a given threshold following the method of \citealt{Robertson2012}. We combine (\S\ref{A3}) the mass--metallicity relation and its redshift evolution \citep{Maiolino2008-hm} with the galaxy mass function \citep{McLeod2021-us} and the galaxy main sequence \citep{Tomczak2014-py}. We find that a metallicity threshold $12+\log(\rm O/H)\le$ 8.6 , represented by the the dashed  blue models in Fig.\ref{fig:rate}, accounts for the derived GRB efficiency evolution. The metallicity thresholds we derive are consistent with the observed GRB host galaxy metallicity \citep{Vergani2015-rr,Palmerio2019-zp}. To further verify this consistency we computed the stellar masses of galaxies with metallicity below the  threshold value found and compare with the observed masses of a complete sample of GRB hosts (\citealt{Vergani2015-rr,Palmerio2019-zp} - black symbols in Fig.\ref{fig:host} left panel). The maximum and average stellar masses (green and blue curves in Fig.\ref{fig:host}-left panel) are shown for $12+\log(\rm O/H)\le$ 8.6 by the dashed lines and are consistent with the measured hosts' masses and their redshift evolution. 

 
 Previous LGRB population studies (e.g. \citealt{Firmani2004,Salvaterra2012-uc,Palmerio2020-sh}) showed that a pure density evolution (PDE) of the GRB formation rate, where the luminosity function is invariant with redshift, and pure luminosity evolution (PLE), where the GRB formation rate is proportional to the CSFR, cannot be distinguished on the basis of the GRB peak flux and luminosity--redshift distributions. Our best model is intermediate between these two extreme scenarios and, indeed, we find both the GRB formation rate evolving with redshift (Fig.\ref{fig:rate}-{\it b}) and a mild evolution  of the characteristic GRB luminosity $\propto (1+z)^{0.64}$ (Tab.\ref{tab:1}).
We at more than $5\sigma$ significance the extreme cases of PLE and PDE. Moreover, the PLE is also excluded by the host galaxy mass distribution (cf red line in Fig.\ref{fig:host}). The finding of an evolution of the LGRB characteristic luminosity $\propto (1+z)^{0.64}$ could hide the change of some properties of the progenitors with redshifts possibly ruled by their metallicity  \citep{Yoon2006-ic} which could induce a flattening of the initial mass function \citep{Fryer2021}. 
This possibility can be tested by an increasing fraction of GRBs detected at high redshifts as envisaged by the 
next generation missions optimised to this purpose (e.g. the Transient and High Energy Sky and Early Universe Surveyor - THESEUS \citealt{Amati2021} or {\it Gamow Explorer} - \citealt{White2020}).  


Our population model, which implements the jet opening angle of GRBs and accounts for the observed properties as a function of the viewing angle, allows us to estimate the fraction of GRBs detected by \fermi/GBM and \swift/BAT which are observed off--axis, i.e. \view$>$\jet. Fig.\ref{fig:host} (right panel) shows the expected cumulative rate as a function of redshifts of \fermi/GBM (blue filled stripe) and \swift/BAT (orange filled stripe). The hatched curves show the detection rates of off--axis events according to our population. We estimate that $\sim$10\% of LGRBs detected by \fermi and \swift at $z<0.5$, corresponding to a rate 0.2--0.7 yr$^{-1}$ respectively, should be observed off--axis. This fraction reduces to less than 3\% at $z<2$. In this work we adopted a uniform jet structure consistently with what suggested by the study of the luminosity function of long GRBs \citep{Pescalli2015-bm}. Therefore, our results should be considered as a conservative lower limit since a universal structured jet should produce a larger fractions (e.g. \citealt{Salafia2015-dj}) of detected off-axis events.

\acknowledgments
We acknowledge support by ASI-INAF agreement n. 2018-29-HH-0. We acknowledge PRIN-MIUR 2017 (grant 20179ZF5KS) and the "Figaro" premiale project (1.05.06.13). We thank S. Vergani, G. Ghisellini and O. S. Salafia for constructive discussions. We acknowledge L. Amati and N. White and the respective THESEUS and {\it Gamow Explorer} collaborations for stimulating interactions over the past years. 




\appendix

\section{Appendix information}


\subsection{Consistency checks}\label{A2}

\begin{figure*}
    \centering
    \includegraphics{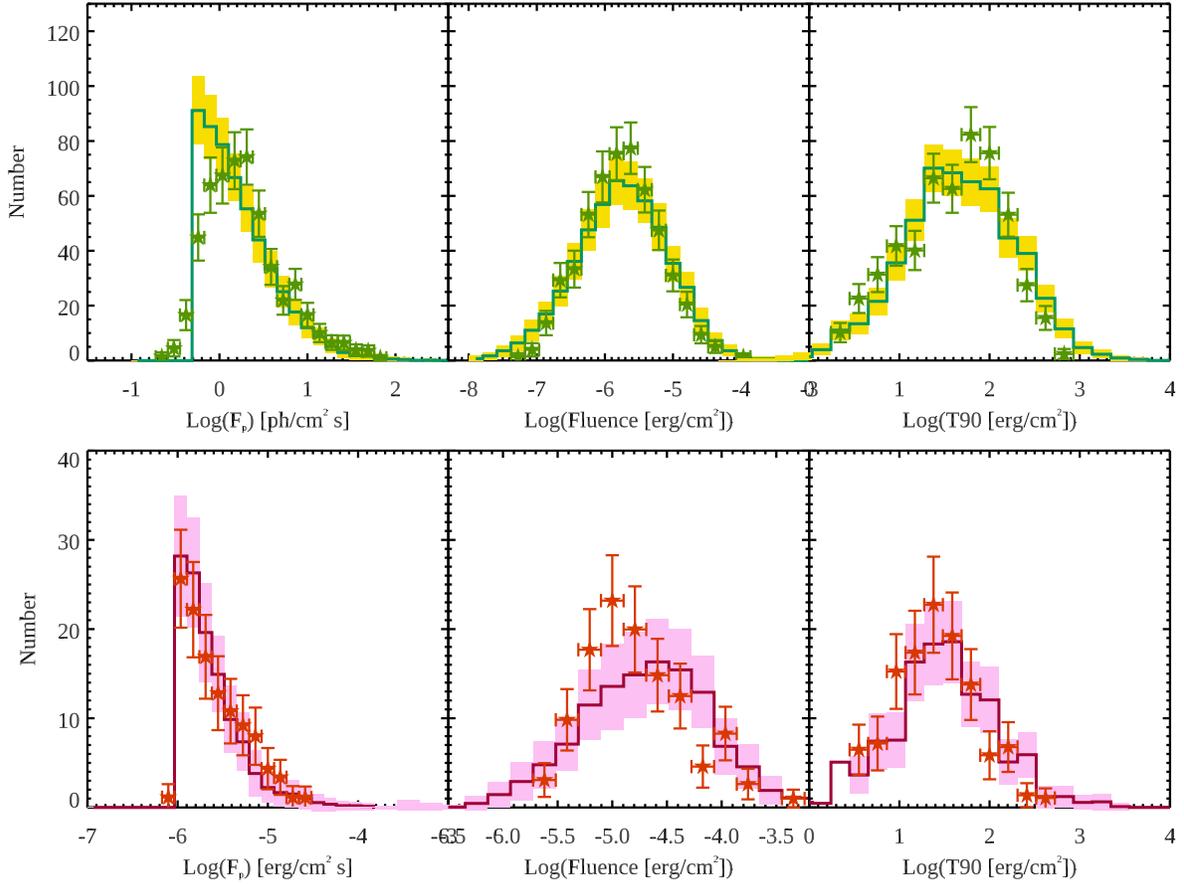}
    \caption{Consistency checks of the population model with the distributions of the peak flux, fluence and duration of GRBs detected by the Swift/BAT satellite (top row) and Beppo--SAX (bottom row). Swift/BAT GRBs (green symbols) are obtained selecting GRBs with a peak flux (15--150 keV) $\ge$0.5 ph cm$^{-2}$ s$^{-1}$ while Beppo--SAX GRBs (pink symbols) are selected with peak flux (40--700 keV) $\ge$ 10$^{-6}$ erg cm$^{-2}$ s$^{-1}$. Model is shown by the solid lines and its 1$\sigma$ uncertainty (described as in Fig.\ref{fig:constr}) by the solid filled histogram.}
    \label{fig:consistency}
\end{figure*}

In order to verify the consistency of our model with GRB samples detected by other instruments we considered the full \swift/BAT and BeppoSAX/GRBM \citep{Frontera2008-el} samples. We applied a peak flux cut extracting 536 \swift/BAT GRBs with a 15-150 keV peak flux $F_{p}\ge 0.5$ phot cm$^{-2}$ s$^{-1}$. Their observer frame distributions are shown by the green asterisks in the top panels of Fig.\ref{fig:consistency}. For Beppo/SAX we considered the 118 GRBs with peak flux, integrated in the 40--700 keV energy range, $F_{p}\ge 10^{-6}$ erg cm$^{-2}$ s$^{-1}$. The BeppoSAX distributions are shown by the pink asterisks in the bottom panels of Fig.\ref{fig:consistency}. The population model is shown by the solid histogram and shaded regions. 

\subsection{Schechter luminosity function}\label{AS}

\begin{figure}
    \centering
    \includegraphics{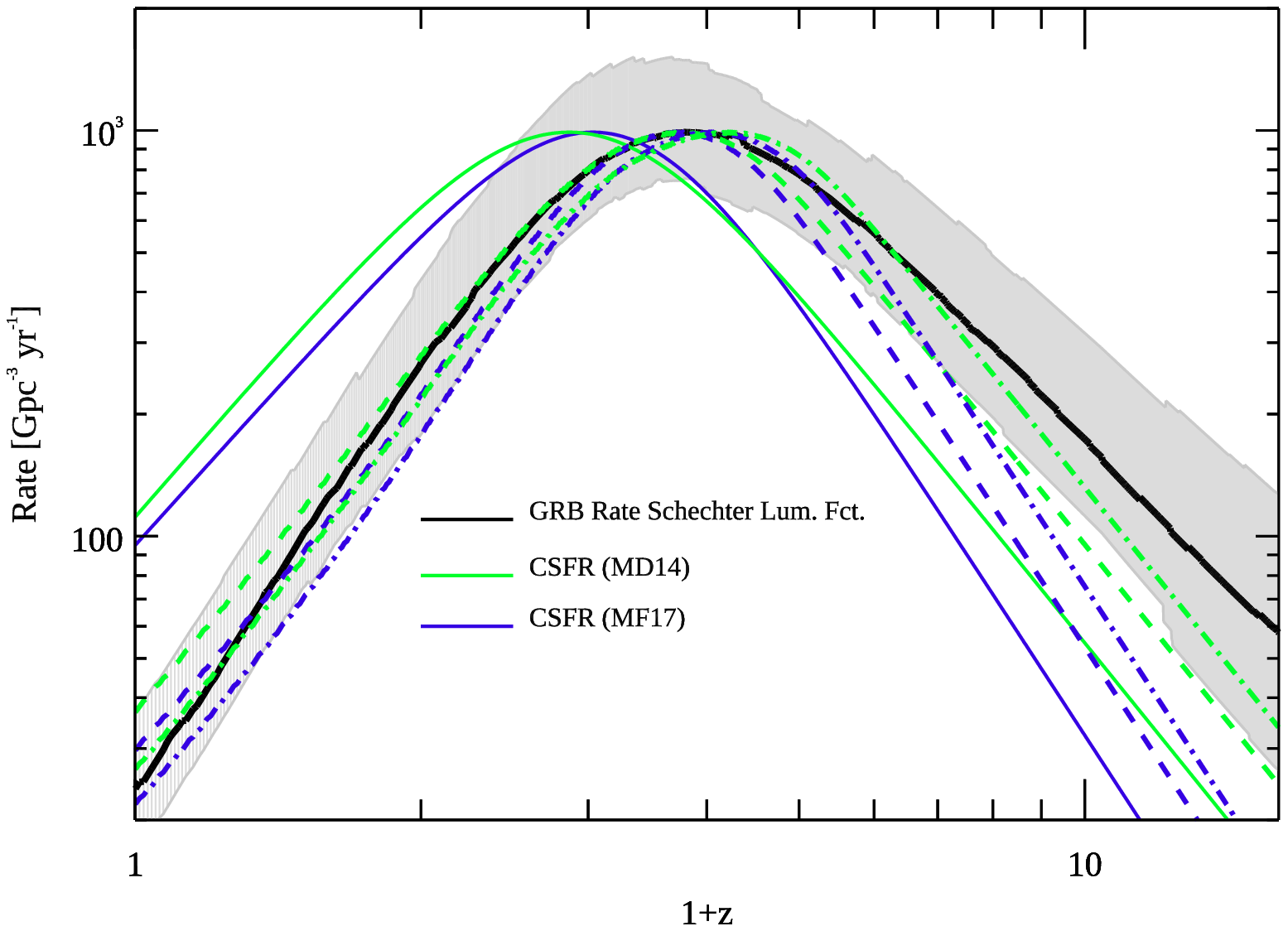}
    \caption{ GRB formation rate (black solid line) and its 1$\sigma$ uncertainty (shaded region) obtained with the model described in \S1 but assuming a Schechter luminosity function instead of Eq.\ref{LF}. The solid blue line shows CSFR as derived by \citealt{Madau2017-hz} and the solid green line that obtained by \citealt{Madau2014-wg}. The dashed (dot--dashed) lines are the CSFR functions modified by a metallicity threshold 12+$\log\rm(O/H)\le$8.6 (8.4) respectively (\S\ref{A3}). Blue and green model curves have been normalized to the peak value of the GRB formation rate.}
    \label{fig:schechter}
\end{figure}

We considered the possibility that the GRB luminosity function is described by a Schechter function \citep{Schechter1976}, namely a powerlaw with a high--luminosity exponential cutoff. This model has two free parameters, the slope of the powerlaw $\alpha$ and the cutoff luminosity $L_c$. We fix the minimum luminosity of this function to $10^{47}$ erg s$^{-1}$. With this model we can reproduce all the observational constraints described in \S3. We obtain $\alpha=1.15\pm0.15$ and 
$\log(L_c)=53.21\pm0.28$. 
The resulting GRB formation rate is shown in Fig.\ref{fig:schechter} by the black line (and its 1$\sigma$ uncertainty - shaded region). For comparison the CSFR of \citealt{Madau2017-hz} (blue line) and of \citealt{Madau2014-wg} (green line) are shown. The modified CSFR obtained with the metallicity threshold cuts at 12+log(O/H)$\le$8.6 and 8.4 are shown by the dashed and dot--dashed lines respectively.  The GRB formation rate obtained under the Schechter assumption is consistent within their uncertainties with that obtained with a broken powerlaw luminosity function though slightly shallower at high redshifts. 

\subsection{Metalliticy bias model}
\label{A3}

Following the formalism of \citealt{Robertson2012}, we estimate the fraction of star formation occurring in galaxies with metallicity below a given threshold as: 
\begin{equation}
    f(z)=\frac{\int_{M_{\star,\rm min}}^{M_{\star,\rm th}(z)} R_{\star}(M,z)\cdot \Upsilon(M,z) dM}{\int_{M_{\star,\rm min}}^{M_{\star,\rm max}} R_{\star}(M,z)\cdot \Upsilon(M,z) dM}
\end{equation}
where $R_{\star}(M,z)$ is the star formation rate -- stellar mass relation ($M$ is the galaxy stellar mass in units of $M_{\odot}$)  and  $\Upsilon(M,z)$ is the galaxy mass function. Both these functions evolve with the redshift $z$. For $R_{\star}(M,z)$ we assume the relation of \citealt{Tomczak2014-py} and for $\Upsilon(M,z)$ we consider the recently reported double Schecter galaxy mass function of \citealt{McLeod2021-us}. $M_{\star,\rm th}(z)$ is the maximum mass corresponding the the metallicity threshold: 
\begin{equation}
    \log M_{\star,\rm th}(z)=\left[\frac{K(z)-\left[12+\log(O/H)\right]}{0.0864}\right]^{1/2}+\log M_{0}(z)-0.04
\end{equation}
obtained by inverting Eq.(2) of \citealt{Maiolino2008-hm}. The values of $K(z)$ and $M_0(z)$ are interpolated from those reported in Tab.5 of their paper.  We also corrected (through the 0.04 term) for the different assumption of a Chabrier initial mass function in host galaxy mass estimates shown in Fig.\ref{fig:host}.

\bibliography{Paper_GGRS}{}
\bibliographystyle{aasjournal}




\end{document}